# Recent Advances in Fibrosis and Scar Segmentation from Cardiac MRI: A State-of-the-Art Review and Future Perspectives

**Yinzhe Wu[1,2*], Zeyu Tang[2], Binghuan Li[2], David Firmin[1,3], Guang Yang[1,3*]**

[1] National Heart and Lung Institute, Faculty of Medicine, Imperial College London, London, UK

[2] Department of Bioengineering, Faculty of Engineering, Imperial College London, London, UK

[3] Cardiovascular Biomedical Research Unit, Royal Brompton Hospital, London, UK

**\* Correspondence:**

Yinzhe Wu and Guang Yang

yinzhe.wu18@imperial.ac.uk and g.yang@imperial.ac.uk

## Abstract

Segmentation of cardiac fibrosis and scar are essential for clinical diagnosis and can provide invaluable guidance for the treatment of cardiac diseases. Late Gadolinium enhancement (LGE) cardiovascular magnetic resonance (CMR) has been successful for its efficacy in guiding the clinical diagnosis and treatment reliably. For LGE CMR, many methods have demonstrated success in accurately segmenting scarring regions. Co-registration with other non-contrast-agent (non-CA) modalities, balanced steady-state free precession (bSSFP) and cine magnetic resonance imaging (MRI) for example, can further enhance the efficacy of automated segmentation of cardiac anatomies. Many conventional methods have been proposed to provide automated or semi-automated segmentation of scars. With the development of deep learning in recent years, we can also see more advanced methods that are more efficient in providing more accurate segmentations. This paper conducts a state-of-the-art review of conventional and current state-of-the-art approaches utilising different modalities for accurate cardiac fibrosis and scar segmentation.

## 1    Introduction

Necrosis regions found on the heart (including left atrium (LA) pre-ablation fibrosis, LA post-ablation scar and left ventricle (LV) infarction), depending on the location and size, can have various implications on the cardiac conditions of the patients. For example, ventricular scars can be signs of earlier episodes of myocardial infarction (MI) [1]–[4]. Locating and quantifying the fibrosis and scars have also been demonstrated as a valuable tool for the treatment stratification of patients with atrial fibrillation (AF) [5], [6] or ventricular tachycardia [7] and provide guidance information for the surgical or ablation based procedures [8]. Imaging of post-ablation scars may also give valuable information on treatment outcomes [9], [10].

Cardiovascular magnetic resonance (CMR) has been one of the modern imaging techniques, which are widely used for qualitative and quantitative evaluation of cardiac conditions and to support diagnosis, monitoring disease progression and treatment planning [11]. In particular, Late Gadolinium enhancement (LGE) CMR has been an emerging technique for locating and quantifying regions of fibrosis and scars across the LA and the LV [10], [12]–[15]. LGE CMR has also been shown to improve ablation strategy planning, treatment stratification and prognosis by pre-ablation fibrosis quantification via clinical validations [16]. It also enabled computationally guided and personalised targeted ablation in treating AF in clinical practices [17].

Many algorithms have been developed in the segmentation of cardiac scarring regions, and a few challenges have benchmarked some of the high-performing methodologies (**Table 1**). Among these, 2-SD (standard deviation) has been advocated by the official guidelines [18], while the full width at half maximum (FWHM) technique has been advocated as the most reproducible method to segment ventricular scars [19] (See Section 3.2 for descriptions of 2-SD and FWHM methods). As these algorithms are usually based on successful segmentation of the corresponding anatomical regions beforehand as an accurate initialisation, there has also been rising attention to the automated segmentation of LA and LV anatomy from the LGE CMR images (**Table 1**).

With the development of artificial intelligence techniques, we can observe a rising number of various deep learning models using convolutional neural networks (e.g., fully connected neural network (FCNN) [20] and U-Net [21]), which have demonstrated encouraging results in segmentations of cardiac substructures in recent years [22]. It has also been found that deep learning can be directly applied to scar segmentation as a fully automated end-to-end solution for the input LGE CMR images. With co-registration of different modalities together and deep learning based transfer learning, the combination of LGE CMR with other CMR imaging modalities (e.g., balanced steady-state free precession (bSSFP)) may further improve the efficacy and efficiency of the segmentation results.

The use of Gadolinium-based contrast agent (GBCA) has led to concerns over the patient's safety, particularly for the patient with renal impairments [23]. With deep learning based methods, cardiac scarring regions can now be localised and quantified in non-Gadolinium enhanced CMR images, i.e., without GBCA injections [24].

As all pre-2016 and pre-2013 cardiac scarring segmentation have been carefully benchmarked and summarised by Karim et al. [25], [26], this paper instead focuses on the survey of the post-2016 methodologies in fibrosis and scars delineation and segmentation of the LA and LV anatomy from LGE CMR images. This study also discusses the potential use of modalities other than LGE CMR in locating and quantifying the scars.

## 1.1   Search Criteria

To identify related contributions, search engines like Scopus and Google Scholar were queried for papers on or after 01 Jan 2016 containing ("atrial" OR "ventricular") and ("cardiac") and ("segmentation") with or without ("scar") in title or abstract. Papers that do not primarily focus on the segmentation of cardiac scar or scar-related cardiac anatomy were excluded. Each paper was reviewed and agreed upon by at least two of us (Y.W., Z.T., B.L.) before inclusion. We found 4,384 papers from the search engines and shortlisted 110 of them following the criterion above (**Figure 1**). After full-text screening for their relevances to the topic, we eventually included 47 of them into the study. The last update to the included papers was on 13 May 2021.

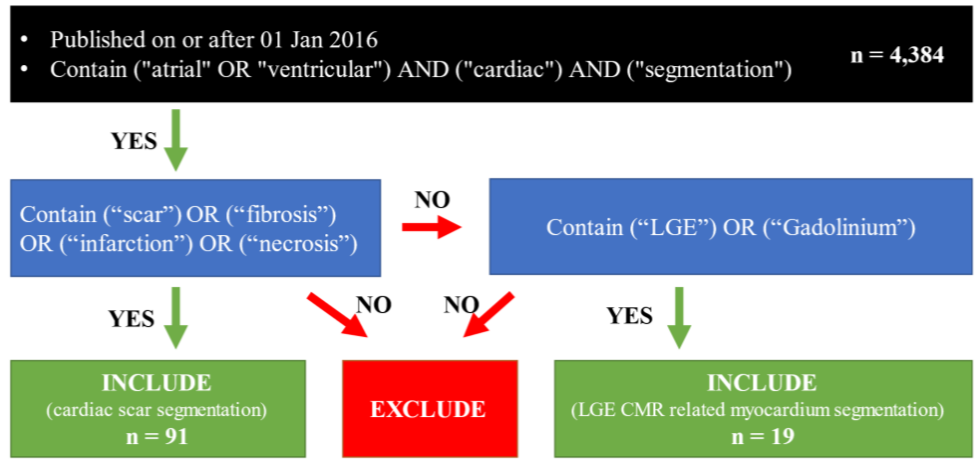

**Figure 1** Flowchart to demonstrate the search criterion.

## 2    Imaging Modalities

### 2.1    LGE CMR

Fibrosis found in LA are signs of atrial structural remodelling and can be considered as a major risk factor in the progression of the atrial fibrillation (AF) [5], [6], where identification of scarring and fibrosis region in LA has been crucial for diagnosis, prognosis and treatment planning. Native pre-ablation fibrosis can be a sign of AF recurrence [13], and post-ablation detection of ablation induced scars can facilitate the identification of post-ablation ablation line gaps, which is the main reason for ablation failures [9], [10]. In contrast to the traditional method of the electro-anatomical mapping (EAM) system, which is an invasive technique in localisation of the atrial scar and the fibrosis with suboptimal accuracy [27], [28], LGE CMR enables the atrial scarring and fibrosis regions to be localised and quantified non-invasively without ionising radiation, where LGE CMR employs the slow washout kinetics of Gadolinium in these regions to highlight these scarring and fibrosis regions [10], [12]–[15].

In addition to the atrium, LGE CMR has also been considered as a gold-standard modality for assessment and quantification of the scarring regions in the left ventricle [29]–[32], where fibrotic and scarring regions found can be considered as a sign for the earlier or current episode of the MI [1]–[4]. In addition to MI, with growing prognostic evidence, LGE has been demonstrated for the identification of scarring regions in cardiomyopathy, inflammatory and infiltrative conditions [33]–[36].

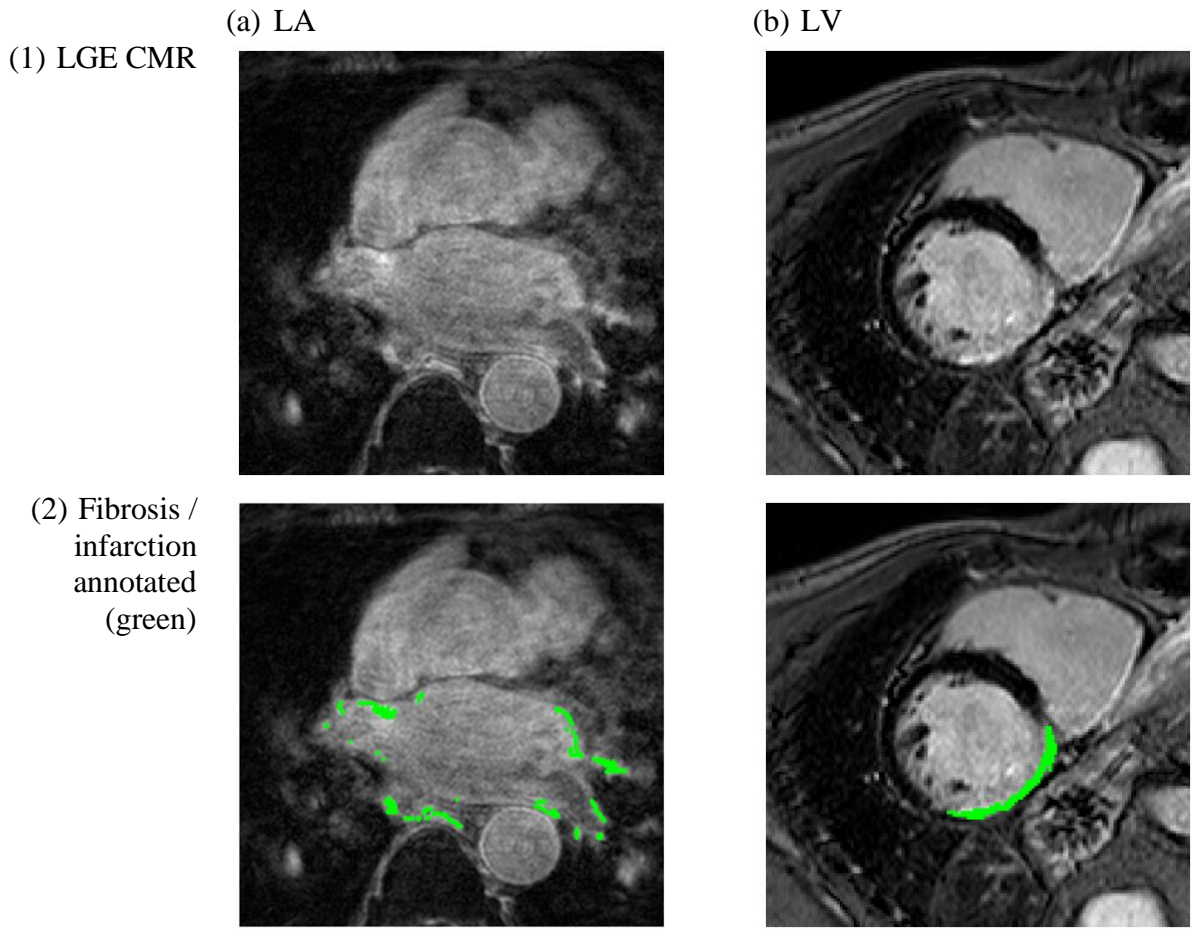

**Figure 2** Examples of LGE CMR images acquired at (a) LA and (b) LV, with the fibrosis/infarction regions highlighted in green. By comparing (a)(2) and (b)(2), we can see the fibrosis region in LA is rather more discrete and thinner compared to LV infarction, making LA fibrosis regions more difficult to be accurately fully localised and quantified.

Image source: (a) was extracted from pre-ablation CMR images in ISBI 2013 cDERMIS dataset (http://www.cardiacatlas.org/challenges/left-atrium-fibrosis-and-scar-segmentation-challenge/). (b) was extracted from MICCAI 2012 Ventricular Infarct Segmentation challenge dataset (http://www.cardiacatlas.org/challenges/ventricular-infarct-segmentation/).

However, the LGE CMR modality often suffers from poor image quality, which may be due to residual respiratory motion, variability in the heart rate and gadolinium wash-out during the currently long acquisition time [37]. Particularly for the left atrium, the spatial resolution of LGE CMR images is limited [38], considering the thin transmural thickness of the atrial wall (mean = 2.2 - 2.5 mm [39]) (**Figure 2**). The variable anatomical morphological shapes of the LA and pulmonary veins (PV) also impose an additional challenge to the LGE CMR segmentations. To improve the visualisation of these necrosis regions, we can see a successful attempt by maximum intensity projection (MIP) to enhance intensities on post-ablation LA LGE CMR [40]. Moreover, some irrelevant cardiac substructures may be highlighted in LGE CMR images as well, in addition to the scarring and fibrosis regions. These may be due to, for example, the navigator beam artefact, which is often seen near the right PV, Gadolinium uptake by the aortic wall and valves and confounded enhancement in the spine, oesophagus, etc.[25],

[37] As a result, these can lead to a poor result in the delineation of LA and LV necrosis regions and even a significant amount of false positives in segmentations of these structures and regions.

In addition, although LGE CMR has been a success as the gold standard reference technique for AF and MI, including LGE in MRI significantly extends the scanning time, and there have been increasingly growing concerns regarding the safety of the Gadolinium based contrast agent used, particularly for the patient with renal impairments [23].

### 2.1.1 LGE CMR With Other Modalities

In addition to LGE MRI, which could highlight the scarring regions, segmentation of the anatomy and scarring regions can also utilise other modalities (**Figure 3**) to further improve the accuracy if applied with LGR CMR by co-registering different modalities together [41].

There have been challenges benchmarking a range of algorithms for the cross-modality fusion based segmentation of anatomy, scar and oedema.

1) MS-CMR challenge [42], [43] presented a range of algorithms taking multiple modalities in to further improve the segmentation accuracy of LV myocardium, LV blood cavity and RV.
2) MyoPS challenge [44], [45] presented algorithms to delineate LV myocardium with scarring and oedema.

Other modalities and sequences can include

1) Magnetic resonance angiography (MRA) sequence – to image LA and PV with high contrasts, which has been demonstrated by Tao et al. [46] to improve the error distance in segmenting LA anatomy to within 1.5mm. However, MRA is usually ungated and usually acquired in an inspiratory breath-hold, making anatomy delineated from MRA significantly distorted from LGE CMR.
2) Balanced steady-state free precession (bSSFP) – provides a clear boundary between the myocardium and blood cavity under movements, which is usually respiratory and cardiac gated. It can offer cine CMR with a uniform texture.
3) T2 – high intensities in T2 presents myocardial oedema with high specificity and sensitivity [47], which could present be helpful in segmenting myocardial oedema and scar simultaneously if incorporated with LGE-CMR and bSSFP [48]. Identification of oedema on CMR can help to differentiate between acute and remote myocardial infarction [49]. The presence of oedema in patients without extensive irreversible injury (e.g., scar) can serve as a marker for clinicians to predict the recovery of LV systolic functions [50].

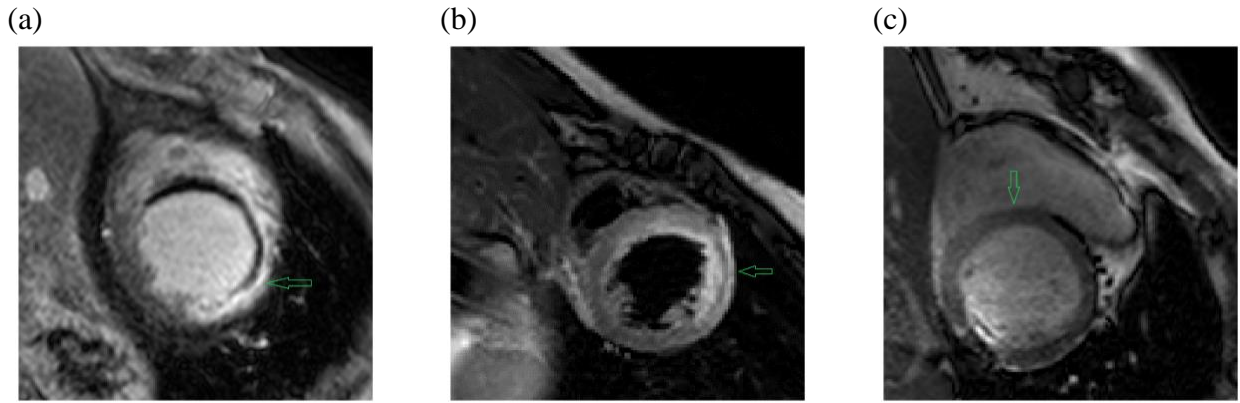

**Figure 3** Example images using different CMR sequences acquired by (a) LGE CMR (b) T2 CMR (c) bSSFP CMR.

As denoted by the green arrows, we can see (a) LGE CMR accentuates the scar tissue by high intensities on the images; (b) T2 CMR accentuates myocardial oedema by high intensities on the image; and (c) bSSFP CMR shows the distinct endo- and epi-cardial boundary of the myocardium clearly on the image.

Image source: (a - c) extracted from the MS-CMR open challenge dataset[42].

## 3 Conventional Methods

Conventionally, a two-stage approach is adopted in the identification and evaluation of fibrotic and scarring tissue – (1) segmentation of the relevant anatomical structure (LA and PV in the case of LA scar segmentation and LV in the case of LV infarction segmentation) and (2) then segmentation of the fibrotic and scarring regions. This two-stage approach is particularly beneficial for LA and PV, as LA and PV are highly morphological variables and relatively small in size. We shall then elaborate on the recent developments of methodologies for each of them.

### 3.1 Segmentation of Anatomical Structures

To differentiate anatomical structures, e.g., LA and LV wall, from others can be difficult in LGE CMR images, as in LGE scarring tissues are significantly enhanced while the signals from the healthy tissues are attenuated [51]. This causes significant challenges for the segmentation of LA, PV and LV anatomical structures.

#### 3.1.1 Why Is Accurate Segmentation of Anatomical Structure Necessary Before Scar Segmentation?

Accurate anatomy (LA or LV wall) is essential as it gives an accurate initialisation for the scar segmentation. Therefore, traditionally, the segmentation of these structures were all done manually.

We could see in the cDEMRIS challenge in ISBI 2012 [25] algorithms with manually initialised LA segmentation showed significantly better performance than others, demonstrating the need for an accurate anatomy segmentation ahead of the scar segmentation along with Rajchl et al. [52]. Moccia et al. also demonstrated that manual and accurate segmentation of the LV wall could improve the deep learning based segmentation of the LV infarction [53].

### 3.1.2 Conventional Methods in Segmenting Anatomical Structures

In the early 21st century, radiologists check between LGE CMR and cine CMR back and forth to delineate the myocardium region. To mimic that, we can see methods in the first decade and early second decade of the century utilising both LGE and cine modalities by, for example, non-rigid registration to achieve high accuracy in segmentation of myocardium over LGE CMR [54]–[57]. However, by doing so, the result may suffer from registration misalignment between LGE and cine modalities and the model may be computationally demanding. As such, from 2014 we can see methods that are less computationally demanding and using LGE modality only [58]–[61].

Conventional methods in medical image segmentation usually have limited efficacy. Representative methods are summarised in **Table 3**, which mainly include the following methodologies.

1) random forest [61]
2) image registration [59]
3) Markov random field (MRF) model [58]
4) atlas-based modelling with active contour model [60]
5) principal component analysis (PCA) technique [59]

For LA, in particular, methods involving pre-defined shape priors [62], [63], although one of them reported a relatively high Dice score (79%) [63], often suffer from error distance more than 1-2mm required [64] under the clinical setting considering the thin LA wall [65].

### 3.2    Segmentation of Scarring Regions

Upon successful segmentation of the anatomy, the scarring regions can be identified by a range of approaches. Mainly these approaches can be divided into the following categories: threshold based methods, classification methods, or the combination of both.

### 3.2.1 Fixed Threshold Based Methods (n-SD and FWHM)

Traditionally, the scarring regions can be detected as they are accentuated in LGE CMR. Among a range of conventional techniques, 2-SD has been advocated by official guidelines [18], while the full width at half maximum (FWHM) technique has been advocated as the most reproducible method to segment ventricular scars [19].

2-SD and FWHM are both fixed threshold methods in segmenting the scarring region, where pixels with intensities above a fixed threshold would be labelled as the scar. 2-SD or even n-SD methods define such threshold as the sum of the mean and two or n standard deviations of signal intensities in a remote reference region, whereas FWHM defines such threshold as the half of the maximum signal intensity within the scar.

Karim et al. [26] evaluated 2, 3, 4, 5, 6 -SD and FWHM methods on a public human LV infarct dataset and showed that FWHM superseded all n-SD methods tested by its Dice Scores and that the Dice Scores went slightly higher with the threshold rising from 2 to 6 -SD. However, it is not the case when Karim et al. [25] evaluated 2, 3, 4 -SD and FWHM on a public human LA fibrosis/scar dataset. For pre-ablation LA fibrosis, FWHM performed much worse than all n-SD methods tested. For post-ablation LA scar, FWHM gave similar Dice Scores as 2-SD's with 3, 4, 6 -SD methods' Dice Scores much lower than these two.

However, these fixed-threshold techniques, including n-SD and FWHM, are unlikely to handle variations well [13]. The variations can come from two sources – scar itself and external circumstances. Scars are highly variable in their morphology and their brightness distribution on LGE CMR. Varied external factors including resolution, contrast, signal-to-noise ratio (SNR), inversion time and surface coil intensity variation can also adversely impact the accuracy of the segmentation. This is particularly the case for pulmonary veins, which are highly morphological variables.

### 3.2.2 Adaptive Conventional Methods

An LV scar segmentation challenge [26] organised in MICCAI 2012 and LA scar segmentation [25] challenge organised in ISBI 2013 carefully benchmarked and summarised the majority of the pre-2013 conventional methods. In the LV segmentation challenge in 2012, it showed all of the algorithms benchmarked did not exhibit superiority against FWHM, although they did perform better than n-SD methods.

### 3.2.2.1 Adaptive Thresholding Based Methods

Conventional threshold based approaches are summarised in **Table 4(A)**, which mainly include the following methodologies.

1) Otsu thresholding [66], [67]
2) histogram analysis [25]
3) hysteresis thresholding [25]
4) constrained watershed segmentation [68]

### 3.2.2.2 Classification based methods

In addition, conventional classification approaches are summarised in **Table 4(B)**, which mainly include the following methodologies.

1) k-means clustering [25]
2) graph cuts [25], [69]
3) active contour with EM-fitting [25]
4) Simple linear iterative clustering (SLIC) and support vector machine [70].
5) random forest classification [71]

## 4  Deep Learning Based Methods

Deep learning based methods are constructed from deep artificial neural networks. In this section, we will briefly introduce the common types of artificial neural networks (ANNs) and then focus on their variants targeting cardiac anatomy and scar segmentations. The authors would also like to recommend interested readers to refer to Goodfellow et al. [72] for more detailed explanations and mathematical illustrations of these networks and Chen et al. [22] for more thorough demonstrations of these networks in general cardiac imaging analysis.

### 4.1  Neural Networks of Deep Learning in Image Analysis

Convoluted neural networks (CNNs), particularly fully convoluted neural networks (FCNNs), have demonstrated success in delineating anatomical structures in medical images [73], especially in cardiac MR [22]. Successful examples include ResNet [20], U-Net [21] and etc. U-Net [21], in particular, has been known for its ability to gather latent information in medical image analysis and thus to gain better performance in segmentation, which has become the most popular CNN backbone architecture, especially after demonstrating success in the ISBI cell tracking challenge in 2015.

The recurrent neural network (RNN) is another type of ANNs. The RNN is rather more useful in processing sequential data, as it could 'memorise' past data and utilise its 'memory' to assist with its current prediction. Widely used structures of RNNs include long-short-term memory (LSTM) [74] and gated recurrent unit (GRU) [75].

Autoencoders (AEs) are also a type of ANNs, which are able to learn latent features of imaging data. Unlike CNNs and RNNs, AEs learn these features without supervision. With latent features gathered by Aes, it could be used to guide the segmentation of medical images [76], [77].

Generative Adversarial Networks (GANs) was initially proposed for image synthesis [78]. With its two-player model structure (a generator network to give a synthesized image and a discriminator network to try to differentiate that synthesized image from a true image), the model can enhance the resolution of the synthesized image by adversarial training. The GAN could also be used for segmentation, where its discriminator network would rather attempt to see if the output label is in an anatomically plausible shape [79].

## 4.2 Segmentation of Anatomical Structures

### 4.2.1 Why Use Deep Learning in the Anatomical Structure Segmentation?

There are a few challenges recently organised to benchmark the new methodologies proposed for the cardiac anatomy segmentation – 2018 LA Segmentation Challenge in MICCAI 2018 (LASC'18) [64] for LA, MS-CMR [42], [43] in MICCAI 2019, and MyoPS 2020 [44], [45] in MICCAI 2020 for LV. With the recent development in deep learning, we can observe a range of methodologies developed for LA and LV segmentation in LGE CMR [80].

In particular, in LASC'18, all deep learning methods had their mean surface distance in LA wall segmentation below 1.7mm, with the minimum mean value of 0.748mm. This demonstrated the efficacy of the deep learning based methods by the surface distance, which is required to be less than 1-2mm under the clinical setting [65].

### 4.2.2 Deep Learning Methodologies in the Anatomical Structure Segmentation

Successful networks demonstrating success in delineating anatomical structures include VGG-net [81], U-Net [82], and V-Net [83].To further exploit the information on the z-axis, LSTM and its variants [84], [85] and dilated residual learning blocks [85] can be introduced to the widely used U-Net.

On top of the U-Net, Xiong et al. proposed a dual path U-Net variant [86], which is demonstrated to have the best Dice Score (0.942) followed by VGGNet (0.864) in their benchmarking of a range of popular CNNs including the original U-Net and one non-deep-learning based method [63] in LA segmentation. Multi-view learning, incorporating axial, sagittal and coronal views together, gave superior performance compared to models based on one view only [87].

On the contrary, further research showed that structural variations in U-Net are unlikely to cause a significant improvement of its performance in LA segmentation from LGE CMR [88], and that deep supervision and attention blocks are unlikely to further improve LA segmentation performance either [89].

In addition to these supervised learning based methods, Chen et al. proposed a feature-matching based semi-supervised learning technique to further improve the segmentation efficacy [90].

All the methods discussed above are summarised in **Table 5**.

### 4.3 Segmentation of Scarring Regions

We can observe a range of deep learning based methods in segmenting scars (**Table 6**).

### 4.3.1 LA Scar Segmentation Models

For LA (**Table 6(A)**), Yang et al. proposed a deep learning based method using Stacked Sparse Auto-Encoders to delineate the LA fibrosis region, which is based on accurate anatomical structure delineation [91]. Li et al. proposed a graph-cuts framework based on multi-scale CNN to further incorporate local and global texture information of the images [92].

### 4.3.2 LV Scar Segmentation Models

For LV (**Table 6(B)**), E-Net [53] and FCNN [93] were demonstrated with high accuracy if with manually segmented LV walls. Then, multi-view U-Net has also been developed in segmenting the scar in a cascaded way [94].

### 4.4 End-to-End Automated Fibrosis and Scar Segmentation

### 4.4.1 Development of End-To-End Scar Segmentation Models Instead of Staged Segmentation Networks

With the development of deep learning, the models can extract further latent information from the LGE CMR images and segment the scar directly from LGE CMR images without acquiring accurate segmentation of the relevant cardiac anatomical structures (e.g., LA wall) in advance while maintaining the accuracy. There has also been a range of methods (**Table 7**) that can complete the segmentation of both the anatomy of cardiac chambers and the scar simultaneously (referred to as "two tasks" below). This is particularly the case for LV, where there is much less variability in its anatomical shape.

### 4.4.2 LA End-To-End Scar Segmentation Models

For LA (**Table 7(A)**), due to the thin LA wall, it is particularly difficult to achieve an end-to-end segmentation of scar directly from LGE CMR. A multi-view two task (MVTT) deep learning based method with dilated attention network was proposed to complete the two tasks simultaneously [95][96]. This study also benchmarked a range of popular deep learning networks such as U-Net and V-Net on each of the two tasks. It compared the performance of its network with conventional methods such as 2-SD and k-means to demonstrate the superiority of its network in completing both of the two tasks accurately on both pre-ablation and post-ablation datasets [96]. This study also suggested that 2-SD, k-means and fuzzy c-means methods clearly over-estimated the enhanced LA scar region [96].

Later, with a joint GAN discriminator, Chen et al. were able to further improve the segmentation accuracy by dealing with the significantly unbalanced two LA targets (LA wall and scar) [97] (**Table 2**). In their method, cascaded learning, a widely applied technique in learning labels with unbalanced classes in natural image segmentation [98]–[104], demonstrated superiority in learning.

### 4.4.3 LV End-To-End Scar Segmentation Models

As LV has less variant morphology and greater size, there have been more successful methods demonstrating efficacy and efficiency in scar segmentation (**Table 7(B)**). E-Net [53] and FCNN [93] were the first few networks that demonstrated the ability to segment scar directly from LGE CMR. Although with relatively low Dice scores, they demonstrated that with an accurately segmented myocardium label it could perform better.

Recently, many deep learning methods have been proposed and demonstrated significantly higher efficacy compared to traditional threshold based methods. Zabihollahy et al. developed a CNN based network to classify each pixel by considering small volume patches around that pixel to greatly improve the mean segmentation accuracy in terms of its mean Dice score to 93.63, compared to the mean Dice scores of K-nearest neighbour (KNN) (77.85), FWHM (61.77) and 2SD (48.33) in their private benchmarking [105].

In addition, Fahmy et al. proved that a 3D CNN deep learning based approach could be applied for LV scar segmentation for patients with hypertrophic cardiomyopathy (HCM) via a multicentre multivendor study [106].

Inspired by the two-stage approach, a multi-view cascaded U-Net driving for even higher efficacy in segmentation was developed to cascade the two tasks sequentially while considering sagittal, axial and coronal views [93].

## 4.5 Segment LGE CMR Jointly with Other Modalities

As explained in Section 3.1.2, traditionally, clinicians check both bSSFP cine and LGE MRI modalities to ensure accurate segmentation of the myocardium and then the scar. Therefore, many methods suggested the use of both bSSFP cine and LGE modalities in delineating anatomical structures and scar to mimic that. For LA, it is also known that MRA gives a clear boundary in PV to help with LA wall segmentation. We can see many methods utilising the MRA modality in addition to others to enhance their segmentation accuracy. However, many studies chose bSSFP over MRA, as bSSFP can be acquired in the same phase as LGE CME by cardiac gating. Although MRA provides better resolution, MRA is not cardiac gated and can be difficult and error-prone in co-registration with LGE CMR, causing misalignments in registered images. Additionally, as explained in Section 2.1.1, integration with other modality (T2 for example) may enable more findings from the CMR (oedema for example) in addition to scars.

1. There are few challenges benchmarking a range of algorithms for the cross-modality fusion based segmentation of anatomy, scar and oedema. MS-CMR challenge [42], [43] presented a range of algorithms taking multiple modalities in to further improve the segmentation accuracy of LV myocardium, LV blood cavity and RV. MyoPS challenge [44], [45] presented algorithms to delineate LV myocardium with scarring and oedema.

Common methods to segment anatomy and scar from multiple modalities include

(1) cross-modality style and feature propagation (typically from bSSFP to LGE-MRI) (e.g., multi-atlas label fusion (MAS) [48])
(2) combination of multiple paired sequences and modalities for segmentation by either cross-modality image style transfer (e.g., Cycle-GAN [48] and UNIT style transfer [107], [108]) or multi-input models (e.g., Multi-variable mixture model (MvMM) [41])

(3) A two-stage approach to firstly co-registering anatomical segmentation from one modality to another (typically from bSSFP segmentation to LGE-MRI) and then segment scars based on the co-registered anatomy segmentation [109].

However, respiratory and/or cardiac motion complications between acquisitions of different modalities can still cause errors in registration and possible misalignments.

## 5    Scar Segmentation With Non-contrast-agent (Non-CA) Enhanced Imaging Modality Only

Although LGE CMR has been very successful as the gold standard reference technique for AF and MI, including LGE in an MRI scanning significantly extends the scanning time, and there have been increasingly growing concerns regarding the safety of the Gadolinium based contrast agent used, particularly for the patient with renal impairments [23]. There has been rising attention in exploring methods to segment scars without injecting contrast agents to the patients on non-CA modalities. Non-CA modality based cardiac scar segmentation has been widely demonstrated for LV scars but has not been realised for LA scars.

Dastidar et al. [110] and Liu et al. [111] demonstrated the potential of pre-contrast scar segmentation by comparing the inter-modality manual observations of myocardial infarction regions on LGE CMR and native-T1 mapping without the Gadolinium contrast agents.

### 5.1    Relaxation Time Based Scar Segmentation in T2

T1 and T2 [112] are modalities that are not enhanced by any contrast materials, where relaxation times in MI is longer compared to the healthy myocardium and could be referenced for MI region segmentation reproducibly [113]–[115]. However, the relaxation time is field strength specific [116], [117] and requires the acquisition of images for additional breath holds, which significantly extends the CMR acquisition time.

### 5.2    MRI Feature Tracking

MRI feature tracking is also an approach to differentiate MI induced cardiac wall abnormalities from normal myocardium, which can be acquired as part of a standard CMR scanning examination [118], [119]. However, this technique can only detect and locate the position of MI without quantifying it.

### 5.3    Scar Segmentation in CINE MRI

To further improve scar segmentation on non-contrast enhanced CMR, trained by co-registered LGE and cine MRI modalities, SVM based texture analysis in pre-contrast cine MRI only can discriminate between nonviable, viable and remote segments [120]. Non-contrasted enhanced CMR scar segmentation has also been demonstrated via neighbourhood approximation forests [121], Simple Linear Iterative Clustering (SLIC) [122] based supervoxels [123].

### 5.3.1  Deep Learning Based Scar Segmentation in CINE MRI

With the development of deep learning, a method based on a combination of Long short-term memory (LSTM), recurrent neural network (RNN) and fully convoluted neural network (FCNN) [124] and a GAN based method [125] have been demonstrated accuracy in detecting, locating and quantifying LV scarring regions from non-contrast enhanced CMR images. Zhang et al. proposed a deep learning based framework to greatly improve the efficacy of the segmentation of LV scar on cine MRI with stages of ROI localisation, RNN based motion pattern extraction, and pixel classification by FCNN and assess their network extensively under a clinical setting [24]. Xu et al., on top of the deep learning based workflow, proposed a progressive sequential causal generative adversarial network (GAN) to simultaneously synthesize LGE-equivalent images and multi-class tissue segmentation (including LV blood cavity, LV myocardium and scar region) from cine CMR images [126]. A detailed summary and results of a private benchmarking of all these algorithms can be found in **Table 8**.

## 6    Evaluation Metrics

A range of evaluation metrics can be employed for assessing the results of the segmentation of the anatomy. These include Dice score, sensitivity, specificity, Hausdorff distance (HD) and surface-to-surface distance (STSD).

### 1)   Dice Score

The Dice Score coefficient, DICE, is one of the most widely used evaluation metrics in segmentation accuracy evaluations. It is particularly sensitive to the difference between the ground truth label and the result label.

Given a 3D prediction label tensor, $A$, and 3D ground truth label tensor, $B$, the Dice score can be defined as

$$\text{DICE}(A, B) = \frac{2|A \cap B|}{|A| + |B|} \tag{1}$$

### 2)   Sensitivity

Sensitivity score, also known as *True Positive Rate*, can be adapted to reflect the success of the algorithm for segmenting the foreground (cardiac anatomy) as

$$\text{Sensitivity} = \frac{\text{TP}}{\text{TP} + \text{FN}} \tag{2}$$

where $TP$ stands for true positive and $FN$ stands for false negative.

### 3)   Specificity

Sensitivity score, also known as *True Negative Rate*, reflects the success of the algorithm for segmenting the background as

$$\text{Specificity} = \frac{\text{TN}}{\text{TN} + \text{FP}} \tag{3}$$

where $TN$ stands for true negative and $FP$ stands for false positive.

### 4)   Hausdorff Distance

Hausdorff distance, HD, is an important parameter in evaluating the geometrical characteristics which measures the maximum local distance between the surfaces of the predicted LA volume label tensor, $A$, and the ground truth label tensor, $B$, given by

$$\text{HD}(A, B) = \max_{b \in B} \left\{ \min_{a \in A} \left\{ \sqrt{a^2 - b^2} \right\} \right\} \tag{4}$$

where $a$ and $b$ are all pixels locations within $A$ and $B$.

In practice, the HD is not generally recommended to use it directly since it has a great sensitivity to outliers, and as noises and outliers are quite common in medical image segmentation [127], [128]. However, Huttenlocher et al. proposed a way to handle outliers by defining the HD as the $q^{th}$ quantile of distance instead of the maximum to exclude the outliers [129].

**5) Surface-to-Surface Distance**

Surface-to-surface distance, STSD, measures the average distance error between the surfaces of the predicted LA volume and the ground truth.

$$\text{STSD}(A, B) = \frac{1}{n_A + n_B} \left( \sum_{p=1}^{n_A} \sqrt{p^2 - B^2} + \sum_{p'=1}^{n_B} \sqrt{p'^2 - A^2} \right) \quad (5)$$

where $n_A$ and $n_B$ are the numbers of pixels in A and B respectively. Variables $p$ and $p'$ describe all point between $A$ and $B$.

Zhao et al. showed that the acceptable minimum error distance in the LA wall segmentation should be 1-2mm under the clinical setting considering the thin LA wall [65].

**6) Error of the Anterior-Posterior Diameter of the Anatomical Structure**

The anterior-posterior diameters of LA and LV are widely used as an essential clinical measure in clinical diagnosis and treatments.

The diameters can be estimated by finding the maximum Euclidean distance along the anterior-posterior axis of each CMR scan [130].

**7) Error of Volume of the Anatomical Structure**

The anatomical volumes of LA and LV are widely used as an essential clinical measure in clinical diagnosis and treatments.

The volume of the structure can be found as the sum of positively labelled voxels. Given the volume of the predicted anatomical structure, $V_A$, and the volume of the ground truth, $V_A$, the total volume error can be defined as

$$\delta V = |V_A - V_B| \quad (6)$$

**8) Scar Volume Percentage**

In addition to the ones mentioned above, scar segmentation also employs a scar volume based metric in assessing the segmentation result, which is much more widely used as the quantification of scar is important for clinical use. They calculate the volumetric percentages of the scarring regions and compare them across the predicted and the ground truth labels.

The scar percentage is defined as the percentage of the volume of the scarring region, $V_{scar}$, relative to the volume of the relevant anatomical wall, $V_{wall}$, (e.g., LA wall) [67].

$$[\%]scar = \frac{V_{scar}}{V_{wall}} \times 100\% \tag{7}$$

## 7 Discussion

### 7.1 Dataset Acquisition

#### 7.1.1 Inter-Observer Variability in the Manual Annotation of Ground Truth Labels

For validation and benchmarking of different methods and training of deep learning based methods, accurate, consistent and reproducible acquisition of ground truth labels is essential.

Validation by employing labels from a single clinician may not be ideal as these labels may exhibit bias and intra-observer variances when the same clinician is asked to repeat their labelling. Thus, it is recommended that we take observations from multiple clinicians and fuse them together.

However, we can see significant inter-observer variances, particularly for LA anatomical segmentation in LGE-MRI where the boundaries of the LA walls are very blurred. Kurzendorfer et al. attempted to compensate for inter-observer variances by additional smoothing but ended up with slight improvement in Dice Scores (+0.04) [59].

It is recommended that the data source reports the inter- and intra- observer variances by employing evaluation metrics such as the Dice Score coefficient. The currently widely used method of label fusing is obtaining a 70% consensus label among multiple annotations, which can be low in their consistency levels. The level of each observer's expertise (novice, medical student, trainee, junior clinician or senior clinician) must also be clearly noted, particularly when multiple observers are involved. It may also be recommended that the observers should all be experienced senior clinicians to maintain the high accuracy and low variance in the manual annotation.

#### 7.1.2 Dataset Sources

Many methods use single-vendor single-centre datasets to validate their methods, which may not demonstrate the ability to generalise the accurate segmentation methodology to centres with CMR machines of different settings and compositions.

There have been some trials assessing the performance of models based on multi-vendor and multi-centre data [106], [131]. However, evaluation based on multi-vendor and multi-centre data with a more significant patient population should be introduced for a more comprehensive unbiased validation, comparison of performances of different methods and assessment for their scalability.

#### 7.1.3 Quantitative Result Reporting

We would like to alert readers that nearly all studies summarised in this study used their own distinct private datasets when reporting results. Biased conclusion may be derived when directly comparing these quantitative metrics across studies. The authors would like to ask all readers to refer to the experimental settings in acquisitions of the datasets stated in their original articles when comparing quantitative results across different studies, instead of only looking at these numbers stated. We would also look forward to a public benchmarking of all these methods as a fairer review of their performances.

Also, the authors would like to ask readers to be cautious when directly comparing Dice Scores reported for the segmentation of the LV infarction than the LA necrosis'. As the LA is much smaller than the LV, an equivalent volume of discrepancy may trigger a more significant reduction in the LA necrosis' Dice Score ratios than the LV infarction's, and the LA necrosis tends to be more challenging to be accurately segmented than the LV infarction explained above.

In addition, the image quality, contrast, class imbalance and other factors of the image data can directly impact the result generated from it and thus the accuracy reported. In particular, the authors would advocate future literature to report (1) scar to blood pool contrast ratios (SC-BP) [69] to show the scar contrast, (2) signal-to-noise ratio (SNR) to show the noise variation along with evaluation metrics in results, so the readers can have a better understanding of the experimental settings before interpreting all the metrics reported quantitatively. These two additional metrics are essential, particularly when it comes to LA scar segmentation, where the scar segmentation is more difficult and where higher SC-BP can give higher Dice Scores in the results generated [69].

## 7.2 Conventional Methods

### 7.2.1 Advantages – Computational Load And Explainability

Obviously, as conventional methods are less demanding on the composition of the computing device, they can be deployed for wider clinical uses more easily. This is an advantage when it comes to the scalability and generalisability of the product, where a standard computer is enough.

Conventional methods are also more explainable than deep learning. The explainability also guarantees easier acceptance from the clinicians, as the product may appear more trustworthy and more reliable.

### 7.2.2 How Reliable Are the Conventional Methods?

#### 7.2.2.1 Fixed Threshold Conventional Methods

Fixed threshold methods may not fit some LGE CMR images, as they are unlikely to handle variations well [13].

Scars are highly variable in their morphology and their brightness distribution on LGE CMR. Some severe LV cardiac scar may appear bright in its surroundings and very dark in its centre, as the centre of the scar is so severely infarcted that very little GBCA carrying perfusion arrives there. N-SD and FWHM, which require the pixel intensity to be more than a certain threshold for that pixel to be recognised as a scar, may not label these dark centres as the scar. Additionally, due to the partial volume effect, fibrotic regions containing both intermingling bundles of fibrotic and viable myocytes will be darker than the complete necrosis region. The low intensity exhibited from such fibrotic regions may be below the fixed threshold set and make these fibrotic regions be falsely recognised as healthy myocardium.

Varied external factors including resolution, contrast, signal-to-noise ratio (SNR), inversion time and surface coil intensity variation can also adversely affect the accuracy of the scar segmentation. LGE CMR modality often suffers from poor image quality, which may be due to residual respiratory motion, variability in the heart rate and gadolinium wash-out during the currently long acquisition time [37]. Considering the thin transmural thickness of the atrial wall (mean = 2.2 - 2.5 mm [39]) (**Figure 2**), the spatial resolution of LGE CMR images is relatively limited, particularly for the left atrium [38]. The variable anatomical morphological shapes of pulmonary veins (PV) also impose an additional challenge to the LGE CMR segmentations. In addition, some uninterested cardiac substructures may be highlighted in LGE CMR images as well in addition to the scarring and fibrosis regions. These may be due to the navigator beam artefact (which is often seen near the right PV), Gadolinium uptake by the aortic wall and valves and confounded enhancement in the spine, oesophagus, etc. [25], [37].

### 7.2.2.2 Adaptive Conventional Methods

Although adaptive conventional methods may mitigate adverse impacts from variable scar shapes and varied external factors, adaptive conventional methods can also be affected by sizes, variances and artefacts in testing image data as they utilise prior information learned. Kurzendorfer et al. [59] showed that a particular scar distribution over the myocardium could adversely affect their methods in segmenting endocardial contours. Such vulnerability may be more problematic when it comes to LA anatomical structures, as PV is a very morphological variable and LA walls are much smaller and thinner.

## 7.3 Deep Learning Based Methods

### 7.3.1 How We Could Make the Deep Learning Perform Even Better?

For detailed designs of the deep learning networks, LASC'18 benchmarked [64] a range of U-Net variants in LA wall segmentation from LGE CMR. This challenge, along with other literature for cardiac scar segmentation, demonstrated the following.

1) **Image Sources**
   a) Higher image qualities (as in signal-to-noise ratio) would result in a higher Dice Score, although not statistically significantly linearly related.
   b) In addition, models with contrast normalisation as a pre-processing technique performed significantly better than the ones without using normalisation.
2) **Model Backbone**
   a) CNN based methods delivered better results compared to the atlas based methods.
   b) U-Net based methods outperformed other networks using VGGNet, ResNet etc.
   c) There was no statistical difference between the segmentation performances of the models based on 2D CNNs and models based on 3D CNNs. However, further research showed that 3D CNNs greatly outperformed 2D CNNs with the same model architecture in terms of the Dice scores of their segmentation results [132].
3) **Segment on ROI or the Whole Image?**
   a) Centring LA on ROI as an input to the second sequential model would make the model perform significantly better compared to the model with non-centred ROIs.
   b) Class imbalance induced by significantly big or small ROI size could lead to an adverse effect on the segmentation results in terms of Dice scores
   c) Double sequential CNNs [64], [133]–[135] (one detecting the region of interest first and then the second model performing regional segmentation within the region of interests (ROI) detected) achieved much better results compared to the methods with only one single CNN.
   d) Double sequential 3D CNN outperformed single 2D CNN and single 3D CNN models regarding its Dice scores, surface distance, LA diameter error and LA volume error.
4) **Model Architecture**
   a) Models with residual connections performed significantly better compared to the ones without residual connections.
   b) The use of dropout blocks did not perform significantly better than the one without using dropout.

c) Rectified Linear Unit (ReLU) trained models did not perform significantly better than the Parametric Rectified Linear Unit (PReLU) trained models.

5) **Loss Functions**

a) Dice loss trained models performed significantly better than the cross-entropy trained models.

### 7.3.2 Problems with Deep Learning in Segmentation

### 7.3.2.1 Computational Load

Although we are able to observe much better results generated from deep learning based methods, we can also observe a rise in computational demand from deep learning networks. For deep learning based methods, high-end computer graphics processing units (GPUs) become a necessity when deploying these models, whereas standard computers with CPUs only are sufficient for most of the conventional methods to run. Under a clinical setting, hiring a GPU is not always possible, as it is not part of a standard clinical computing workstation. The requirement of a high-end computer with GPU in deploying a deep learning based method may significantly limit the ability of these methods to scale.

However, if a standard computer was only to infer a deep learning model, the runtime may be a bit long but still falls within the maximum time limit that clinicians can accept (usually a few minutes per slice for models that are not extra complex). Therefore, we can see these models can be deployed and scaled only if they are sophisticatedly trained, as training on the clinician's side, where unlikely they have a GPU, is not usually possible. As the inference time may vary significantly across different models over CPUs and depend on their architectures and complexities, reporting of inference time per slice on a standard computer without a GPU should also be mandatory in addition to the inference time over a GPU.

### 7.3.2.2 Scarcity of Annotated Data

Training datasets with abundant paired labels are essential to the success of deep learning model training. However, there has been a scarcity of labels due to the tedious process of manually annotating the ground truths in medical imaging. In order to mitigate such scarcity in ground truth labels, several methods can be adopted, including the following.

(1) data augmentation,

(2) transfer learning with fine-tuning [136]–[138],

(3) weak and semi-supervised learning [137], [139]–[143],

(4) self-supervised learning [144] and

(5) unsupervised learning [145].

In addition, to mitigate the challenging training process brought by the great data size required to train a scalable network, active learning [146] has been introduced to reduce manual annotation workloads as well as the computational loads.

### 7.3.2.3 Explainability in Deep Learning

Although there has been a wide range of evidence demonstrating the efficacy of deep learning in medical image analysis, the deep learning networks behave more like a 'black box', where its interpretability is poor. It has been shown that these deep learning networks can be attacked by adversarial noises or even just rotation in medical images [147], questioning the reliability and scalability of these deep learning models in assisting diagnosis. For alerting users of these possible failures, segmentation quality scores [148] and confidence maps (e.g., uncertainty maps [149] and attention maps [150]) should be provided to highlight uncertainties in the model prediction.

## 7.4 Non-CA Modality Segmentation: Bye-Bye to Gadolinium?

Although many methods can accurately segment scars on non-CA cine MRI, the impact from different numbers of cardiac phases on cine MRI has not been assessed.

In addition, the binary class of either normal or scar may be too simplistic. Quantification of the so-called "grey-zone", which has been proposed for the clinical implication of ventricular arrhythmia [151], immediately surrounding the ventricular scar may be useful clinically.

It may also worth notice that gadolinium based contrast agent is not only applied for scar imaging but also for assessing myocardial perfusion, which is usually assessed together in LGE CMR, for which additional classification and differentiation of ischemic and remote regions of myocardium would be useful [152]. Liu et al. demonstrated non-Gadolinium contrast adenosine stress and rest T1 Mapping for identification and classification of normal, infarcted, ischemic and remote regions in LV myocardium [153].

We are glad to see a range of algorithms demonstrated for LV scar segmentation in non-contrast enhanced CMR. However, this has not been demonstrated for CMR images of LA, which is more difficult as the LA scarring regions in CMR suffers from greater variances in morphology and relatively lower resolution of CMR, and LA scars can appear in discrete regions (**Figure 2**), which imposes further challenges to the LA scar segmentation without contrast agents.

## 8 Conclusion

This study summarises the recent developments in cardiac scar segmentation, covering a wide range of conventional and deep learning techniques. In particular, we presented and discussed the usefulness of non-LGE modalities in cardiac anatomy and scar segmentation. We then further discussed the recent progress in segmenting the cardiac scarring region from non-contrast-enhanced images. We hope this review can provide a comprehensive understanding of the segmentation methodologies for cardiac scar and fibrosis and increase the awareness of common challenges in these fields that can call for future research and contributions.

## 10    Tables

| Year | Challenge/Dataset | Conference (MICCAI / IBSI etc.) | Modality (data size n) | Target | Pathology |
|------|-------------------|--------------------------------|------------------------|--------|-----------|
| 2012 | LV scar segmentation challenge [26] | MICCAI | LGE MRI (30) | LV scar | MI |
| 2013 | LA scar segmentation challenge [25] | ISBI | LGE MRI (30) | LA scar | AF |
| 2018 | LA segmentation challenge [64] | MICCAI | LGE MRI (150) | LA cavity | AF |
| 2019 | Multi-sequence Cardiac MR Segmentation Challenge (MS-CMR) [42] | MICCAI | LGE MRI, T2 MRI, bSSFP MRI (45, coregistered) | LV blood pool, RV blood pool, LV myocardium | MI |
| 2020 | Myocardial pathology segmentation combining multi-sequence CMR (MyoPS) [44] | MICCAI | LGE MRI, T2 MRI, bSSFP MRI (45, coregistered) | LV blood pool, RV blood pool, LV normal myocardium, LV myocardial oedema, LV myocardial scar | MI |

**Table 1** List of challenges in segmentation of LV and LA anatomy and scar in LGE CMR.

|  | LA and PVs | | LA scar | |
|---|---|---|---|---|
|  | Dice Scores | ASD (mm) | Dice Scores | ASD (mm) |
| 2D U-Net | 0.898±0.034 | 3.81±3.89 | 0.526±0.118 | 1.83±0.891 |
| 3D U-Net | 0.895±0.032 | 2.55±3.14 | 0.508±0.106 | 1.90±0.837 |
| MVTT [96] | 0.902±0.037 | 2.25±1.39 | 0.613±0.131 | 1.39±1.03 |
| JAS-GAN [97] | 0.913±0.027 | 2.24±2.73 | 0.621±0.110 | 1.24±1.04 |

**Table 2** Result of a private benchmarking [97] of different algorithms on the LASC'18 dataset, reported in their mean ± standard deviation. ASD: Average Surface Distance.

| Reference | Modalities | Methodology description | Pros | Cons | Quantitative result (myocardium) | Dataset |
|---|---|---|---|---|---|---|
| Dikici et al., 2004 [54] | LGE MRI, cine MRI | 1) Define LV border – non-rigid registration of cine and LGE MRI<br><br>2) LV pixel classification - SVM | Automatic segmentation of LGE-MRI with CINE-MRI information | No longitudinal axis (LAX) consideration, resulting in inter-slice misalignment;<br><br>Need to register with other modality (CINE MRI) | Average contour pixel location error = 1.54 pixel | Private (LV LGE + cine MRI, n = 45) |
| Ciofolo et al., 2008 [55] | LGE MRI, cine MRI | 2D segmentation with a geometrical template (LGE only) and 3D mesh alignment (LGE + CINE) | Overcome non-homogeneous intensity of the myocardium in LGE infarcted regions | Meshes focus only on features in the SAX slices, no inter-slice consideration and thus inter-slice misalignment;<br><br>Need to register with other modality (CINE MRI) | ASD = 2.2 mm (endocardial), 2.0 mm (epicardial) | Private (LV LGE + cine MRI, n = 27) |
| Wei et al., 2011 [56] | LGE MRI, cine MRI | 1) Affine transformation estimation<br><br>2) non-rigid registration of LGE and cine MRI<br><br>3) myocardial contour generation by simplex mesh geometry | Utilise information better in connecting cine and LGE MRI | No LAX consideration, resulting in inter-slice misalignment;<br><br>Need to register with other modality (CINE MRI) | Mean Dice = 0.8249; ASD = 0.97 pixel (endocardial), 0.93 pixel (epicardial) | Private (LV LGE + cine MRI, n = 10) |

| Wei et al., 2013 [57] | LGE MRI, cine MRI | Translational registration of LGE and cine MRI data; 3D nonrigid deformation of the myocardial meshes by both short axis (SAX) and longitudinal axis (LAX) data | Consistent and robust segmentation; Consider both SAX and LAX data to reduce interslice misalignment | Need to register with other modality (CINE MRI) | Mean Dice = 0.9409; ASD = 0.67 mm (endocardial), 0.69 mm (epicardial) | Private (LV LGE + cine MRI, n = 21) |
|---|---|---|---|---|---|---|
| Albà et al., 2014 [58] | LGE MRI | Slice-by-slice graph cuts (GC) with interslice and shape constraints | Impose morphological constraints that are common across MRI sequences – no need for subject-specific tuning or for user initialsation and generalisable for other sequences (CINE-MRI); Achieve robustness to variations in grey-level appearance and to image inhomogeneities – more robust to the presence of abnormalities; Consider interslice interactions; No need to register with other modality (e.g. Bssfp cine MRI) | Give poorer result when generalised to CINE-MRI (due to many artefacts in the dataset tested) | Mean Dice = 0.81; ASD = 1.83 mm (endocardial), 2.38 mm (epicardial) | Private (LV LGE MRI, n = 20) |
| Tanja Kurzendorfer, Forman, et al., 2017 [59] | LGE MRI | 1) LV localisation – image registration  2) short axis estimation - principal component analysis (PCA) | Fast speed and low computational workload by using simple texture features; Consider image data along the longitudinal axis in | Poor performance in apex and LV outflow tract, poor accuracy in basal regions; Since this method is texture based, the | Mean Dice = 0.92; ASD = 1.35 mm | Private (LV LGE MRI, n = 30) |
| | | 3) endocardial refinement - a minimal cost path search (MCP) in polar space using the edge and scar information<br><br>4) epicardial refinement - by shape and inter-slice smoothness constraints<br><br>5) surface extraction – 3D mesh generation by marching cube algorithm [154] | addition to the short axis, improving inter-slice smoothness and avoid inter-slice shift;<br><br>No need to register with other modality (e.g. Bssfp cine MRI) | distribution of scar and the small size of the atrium adversely affect its performance | | |
| Tanja Kurzendorfer, Brost, et al., 2017 [60] | LGE MRI | 1) LV detection - circular Hough transforms<br><br>2) LV blood pool detection - morphological active contours approach without edges (MACWE)<br><br>3) endocardial boundary extraction - a minimal cost path search (MCP) in polar space using the edge and scar information<br><br>4) epicardial boundary extraction – by edge information while considering endocardial contour extracted | Fast speed and low computational workload by using simple texture features;<br><br>No need to register with other modality (e.g. Bssfp cine MRI) | Poor performance in apex and LV outflow tract, poor accuracy in basal regions;<br><br>Since this method is texture based, distribution of scar adversely affect its performance | Mean Dice = 0.85 (endocardial) , 0.84 (epicardial);<br><br>ASD = 2.54 mm (endocardial), 3.32 mm (epicardial) | Private (LV LGE MRI, n = 26) |
| T Kurzendorfer | LGE MRI | 1) LV detection - circular Hough transforms, Otsu | Fast speed and low computational workload by | Poor performance in apex and LV outflow tract, resulting in poor accuracy in basal | Mean Dice = 0.83 | Private (LV LGE MRI, n=100) |

| | | | | | |
|---|---|---|---|---|---|
| et al., 2017 [61] | | thresholding and circularity measures<br><br>2) ROI detection - morphological active contours approach without edges (MACWE)<br><br>3) endocardial boundary extraction - random forest classifier<br><br>4) epicardial boundary extraction - minimal cost path search to the boundary cost array in polar space | using simple texture features;<br><br>No need to register with other modality (e.g. Bssfp cine MRI) | regions and poor ASD result | (endocardial) , 0.83 (epicardial);<br><br>ASD = 3.55 mm (endocardial), 4.12 mm (epicardial) | |

**Table 3** Summary of representative conventional methodologies for segmentation of myocardium on LGE-MRI.

| Type of method | Reference | Method Description | Pros | Cons | Quantitative result (necrosis) | Dataset |
|---|---|---|---|---|---|---|
| (A) Thresholding | Hennemuth et al, 2008[68] | Histogram analysis with constrained watershed segmentation | Automatic threshold determination; No training (supervision) needed; | Based on fixed models - mismatches occur for some cases | * | Private (LGE MRI, n = 21) |
| | Tao et al, 2010 [67] | Otsu thresholding [66] Refine segmentation – (accept false rejection) connectivity filtering and (reject false acceptance) region growing | Automatic threshold determination; No training (supervision) needed; No specific density model assumed – no overfitting; Region growing technique can be useful for small MI | Connectivity filtering and region growing may not be suitable for discrete LA fibrosis regions | Mean Dice = 0.83 | Private (LV LGE MRI, n = 20) |
| | Cates et al, 2013 (part of Karim et al, 2013 [25]) | Histogram analysis and simple thresholding | Simple and accurate processing | Time consuming (require manual work); Manual variance may be significant for the thin LA wall | Median Dice = 0.42 (pre-ablation); Median Dice = 0.78 (post-ablation) | ISBI cdermis 2013[25] (LA LGE MRI, n = 30 (pre-ablation), 30 (post-ablation)) |
| | Bai et al, 2013 (part of Karim et al, 2013 [25] | Hysteresis thresholding [155] | Coherent segmentation (adjacent faint scar sections can still be segmented) | Fixed parametrised model relying on emprical data | Median Dice = 0.37 (pre-ablation); Median Dice = 0.76 (post-ablation) | ISBI cdermis 2013[25] (LA LGE MRI, n = 30 (pre-ablation), 30 (post-ablation)) |
| (B) Classification | Perry et al, 2013 (part of Karim et al, 2013 [25]) | K-means clustering | Relatively higher performance in pre-ablation fibrosis segmentation result benchmarking; No training (supervision) needed | Cluster number to be determined beforehand; Variance in LA scar segmented | Median Dice = 0.45 (pre-ablation); Median Dice = 0.72 (post-ablation) | ISBI cdermis 2013[25] (LA LGE MRI, n = 30 (pre-ablation), 30 (post-ablation)) |
| | Karim et al, 2013 (part of Karim et al, 2013 [25]) | Markov random fields (MRF) model with graph-cuts | Relatively higher performance in pre-ablation fibrosis result benchmarking; | Require necessary post-processing steps to refine clustering | Median Dice = 0.30 (pre-ablation); Median Dice = 0.78 (post-ablation) | ISBI cdermis 2013[25] (LA LGE MRI, n = 30 (pre-ablation), 30 (post-ablation)) |

| | | | | | |
|---|---|---|---|---|---|
| | Gao et al, 2013 (part of Karim et al, 2013 [25]) | Active contour with expectation-maximisation(EM)-fitting | Counteract region leaking problem in region growing techniques | Fixed number of Gaussian mixtures in model | Median Dice = 0.42 (pre-ablation); Median Dice = 0.78 (post-ablation) | ISBI cdermis 2013[25] (LA LGE MRI, n = 15 (post-ablation)) |
| | Karim et al, 2014 [69] | Graph cuts | Does not requires manual outlining of base-line healthy myocardium | Require additional modality (bssfp) | * | Private (LA LGE + bssfp MRI, n = 15) |
| | Yang et al., 2018 [70] | Simple linear iterative Clustering (SLIC) + support vector machine | Fully automatic scar segmentation; Able to complement minor flaws in manual annotation | Require collection of b-SSFP modality; Supervised learning – need paired manual labels for training | Mean Dice = 0.79 | Private (LA LGE + bssfp MRI, n = 11 (pre-ablation), 26 (post-ablation)) |
| | Kurzendorfer et al., 2018 [71] | Fractal Analysis and Random Forest Classification | Utilise texture information in addition to clustering | Require accurate segmentation of the myocardium | Mean Dice = 0.66 | Private (LV LGE MRI, n = 30) |

**Table 4** Summary of representative conventional methodologies for segmentation of cardiac necrosis region on LGE-MRI.
*Overall quantitative metric for the whole result population was not found. Please refer to the original article for more information of the result.

| Reference | Model backbone | Method description | Pros / cons | Quantitative result (myocardium) | | | Dataset |
|---|---|---|---|---|---|---|---|
| Zabihollahy et al., 2019 [82] | U-Net | Standard U-Net | Fast processing; deep latent network | Mean Dice = 0.8661 | | | Private (LV LGE MRI, n = 24) |
| Zhang et al., 2020 [84] | U-Net | U-Net with bidirectional convolutional LSTM | Process spatial sequential information | Mean Dice = 0.906 | | | LASC'18 [64] (LA LGE MRI, n = 100) |
| Yang et al., 2018 [85] | U-Net | U-Net with multiview sequential learning via convolutional LSTM and dilated residual learning | Process spatial sequential information on all 3 spatial axes | Mean Dice = 0.897 | | | Private (LA LGE MRI, n = 100) |
| Xiong et al., 2019 [86] | FCNN | Dual-path FCNN concerning both local and global view | Mitigate class imbalance; Less input image size - save GPU memory | Dice = 0.942 | Benchmarking (Dice) | | Private (LA LGE MRI, n = 40 (pre-ablation), 70 (post-ablation) |
| | | | | | U-Net [21] | 0.642 | |
| | | | | | Dilated U-Net [156] | 0.687 | |
| | | | | | Vggnet [156] | 0.684 | |
| | | | | | Inception [157] | 0.792 | |
| | | | | | Resnet [158] | 0.804 | |
| | | | | | DCN-8 [159] | 0.558 | |
| | | | | | Deconv-Net [160] | 0.500 | |
| | | | | | Segnet [161] | 0.656 | |
| | | | | | V-Net [83] | 0.696 | |
| | | | | | Deeporgan [162] | 0.632 | |
| | | | | | Zhu et al. [63] | 0.821 | |
| Xiao et al., 2020 [87] | FCNN | 3D FCNN with 3D view fusion | Process spatial information on all 3 spatial axes volumetrically; Greater GPU memory occupied | Dice = 0.912 | | | LASC'18 [64] (LA LGE MRI, n = 100) |
| Chen et al., 2019 [90] | Double-sided FCNN | Semi-supervised learning - discriminative feature learning via double-sided domain adaptation | Achieve a fusion of the feature spaces of labelled data and unlabeled data to achieve semi-supervision | Mean Dice = 0.9078 | | | Private (LA LGE MRI, two-centre, n1 = 175, n2 = 94) |

**Table 5** Summary of representative deep learning based methodologies for segmentation of myocardium on LGE-MRI.
We included the benchmarking quantitative results from Xiong et al. [86] for readers' interests, as they covered nearly all popular deep learning models for general image processing.

| LA / LV | Reference | Model backbone | Model description | Pros / Cons | Quantitative results (necrosis) | Dataset |
|---------|-----------|----------------|-------------------|-------------|--------------------------------|---------|
| (A) LA | Yang et al., 2017 [91] | Auto Encoder | Stacked Sparse Auto-Encoders | Significantly higher accuracy; Misenhancement in valves etc. can cause false positive; Hyper-parameter sensitive | Mean Dice = 0.82 | Private (LA LGE MRI, n = 10 (pre-ablation), 10 (post-ablation)) |
| | Li et al., 2020 [92] | CNN | Graph-cuts framework based on multi-scale CNN | Multi-scale consideration enables both local and global feature extraction; Surface projection mitigate difficulty in accurate LA wall delineation; Require collection of b-SSFP | Mean Dice = 0.898 | Private (LA + bSSFP, LGE MRI, n = 58 (post-ablation)) |
| (B) LV | Moccia et al., 2018 [53] | E-Net | E-Net on manually segmented myocardium region only | Significantly higher accuracy; Require manual intervention in myocardium segmentation | Dice = 0.86 | Private (LV LGE MRI, n = 30) |
| | Moccia et al., 2019 [93] | FCNN | FCNN on manually segmented myocardium region only | Significantly higher accuracy; Require manual intervention in myocardium segmentation | Median Dice = 0.7125 | Private (LV LGE MRI, n = 30) |
| | Zabihollahy et al., 2020 [94] | U-Net | Cascaded multi-view U-Net via majority vote multi-view fusion | Consider sequential spatial information on all three axes | Median Dice = 0.8861 | Private (LV LGE MRI, n = 34) |

**Table 6** Summary of representative deep learning based methodologies for segmentation of cardiac necrosis regions on LGE-MRI.

| LA / LV | Reference | Model backbone | Model description | Pros / Cons | Quantitative results (necrosis) | Dataset |
|---|---|---|---|---|---|---|
| (A) LA | Yang et al., 2020 [96]* | ResNet | Multi-view based dilated attention and residual network with sequential learning via convolutional LSTM | Spatial sequential information processing; Attention network to tackle class imbalance | Mean Dice = 0.8258 | Private (LGE MRI, n = 190 (97 pre- and 93 post- ablation)) |
| | Chen et al., 2021 [97] | GAN | Adaptive attention cascade network for simultaneous estimation of unbalanced targets + joint discriminative network for adversarial regularisation | Inter-cascade adversarial learning paradigm to tackle class imbalance and regularise the output | Mean Dice = 0.946 | Private (LGE MRI, n = 192 (97 pre- and 95 post- ablation)) |
| (B) LV | Moccia et al., 2018 [53] | E-Net | E-Net | Relatively low accuracy; Unable to tackle class imbalance well | Dice = 0.55 | Private (LV LGE MRI, n = 30) |
| | Moccia et al., 2019 [93] | FCNN | FCNN | Relatively low accuracy; Unable to tackle class imbalance well | Median Dice = 0.5400 | Private (LV LGE MRI, n = 30) |
| | Zabihollahy et al., 2019 [105] | CNN | Volume patch based 3D CNN | utilise small volume patches for accurate local view inspection | Mean Dice = 0.9363 | Private (LV LGE MRI, n = 10) |
| | Fahmy et al., 2020 [106] | U-Net | U-Net based 3D CNN | Sub-volume design utilises small volume patches for accurate local view inspection | Mean Dice = 0.54 | Private (LV LGE MRI, multi-vendor n = 1073) |

**Table 7** Summary of representative end-to-end deep learning based methodologies for segmentation of cardiac necrosis regions on LGE-MRI.
* As [96] and [95] reported very similar methodologies, we reported [96] only in this table.

| Reference | Method desciption | Pros / Cons | Private Benchmarking Accuracy (%) [126] (necrosis) | Dataset |
|---|---|---|---|---|
| Xu 2020 [126] | (1) priori coarse tissue mask generation GAN,<br>(2) condition LGE-equivalent image synthesis GAN,<br>(3) fine segmentation GAN | Segment more than just cardiac necrosis - LV blood pool, myocardium and necrosis region; Further improve temporal-spatial learning by a two-stream structure that includes a spatial perceptual pathway, a temporal perceptual pathway, and a multi-attention weighing unit. | 97.13 | Private (SAX cine bSSFP MRI, Xu, 2020 [126], n = 280) |
| Zhang 2019 [24] | (1) LV localisation – ROI detection by CNN<br><br>(2) Motion feature extraction<br>(2.1) global motion feature – dense motion flow estimation<br>(2.2) local motion feature – LSTM-RNN<br><br>(3) infarction discrimination – FCNN | Combine both LSTM-RNN based local motion analysis and dense motion flow estimation based global motion analysis | 95.03 | |
| Xu 2018 [125] | GAN<br>(A) Generator:<br>(A1) LV morphology and kinematic abnormalities - spatio-temporal feature extraction network through 3D successive convolution<br>(A2) complementarity between segmentation and quantification - joint feature learning network for multitask learning;<br><br>(B) Discriminator:<br>(B1) intrinsic pattern between tasks – uses task relatedness network for adversarial learning | Introduce adversarial learning and task relatedness to reduce divergence | 96.77 | |
| Xu 2017 [124] | (1) Heart localisation – FAST R-CNN [163]<br>(2) Motion statistical featuer – LSTM-RNN<br>(3) discriminative layer - FCNN | Combine both ROI based local motion analysis and deep optical flow based global motion analysis | 94.93 | |

| Popescu 2017 [123] | Simple Linear Iterative Clustering (SLIC) based supervoxels [122] | Only radial strain analysed, excluding longitudinal and circumferential strains; K-means clustering used requires an empirical definition of the number of clusters | 86.47 | |
|---|---|---|---|---|
| Bleton 2016 [121] | Neighbourhood approximation forests | Consider myocardial thickness and its temporal variations | 84.39 | |

**Table 8** Summary of representative machine learning/deep learning based scar segmentation in cine MRI for segmentation of cardiac necrosis regions on cine bSSFP MRI.